\documentclass[twocolumn,showkeys,showpacs,preprintnumbers,prd,superscriptaddress,nofootinbib]{revtex4-1}
\usepackage{graphicx}
\usepackage{epsf}
\usepackage{bm}
\usepackage{amsmath}
\usepackage{amsfonts}
\usepackage{amssymb}
\usepackage{epstopdf}
\usepackage{natbib}
\usepackage{hyperref}
\usepackage{color}
\usepackage{verbatim}
\usepackage{multirow}
\usepackage{bm}
\usepackage{hyperref}

\makeatletter\let\expandableinput\@@input\makeatother

\newcommand{\deltadmde}{\ensuremath{\delta{}_{DMDE}}}
\newcommand{\Omegafld}{\ensuremath{\Omega_{\rm 0,fld}}}
\newcommand{\wfld}{\ensuremath{w_{\rm 0,fld}}}

\begin{document}

\title{Late time interacting cosmologies and the Hubble constant tension }

\author{Stefano Gariazzo}
\email{gariazzo@to.infn.it}
\affiliation{Istituto Nazionale di Fisica Nucleare (INFN), Sezione di Torino, Via P. Giuria 1, I-10125 Turin, Italy}

\author{Eleonora Di Valentino}
\email{e.divalentino@sheffield.ac.uk}
\affiliation{School of Mathematics and Statistics, University of Sheffield, Hounsfield Road, Sheffield S3 7RH, United Kingdom}

\author{Olga Mena}
\email{omena@ific.uv.es}
\affiliation{IFIC, Universidad de Valencia-CSIC, 46071, Valencia, Spain}

\author{Rafael C. Nunes}
\email{rafadcnunes@gmail.com}
\affiliation{Instituto de F\'{i}sica, Universidade Federal do Rio Grande do Sul, 91501-970 Porto Alegre RS, Brazil}
\affiliation{Divis\~{a}o de Astrof\'{i}sica, Instituto Nacional de Pesquisas Espaciais, Avenida dos Astronautas 1758, S\~{a}o Jos\'{e} dos Campos, 12227-010, S\~{a}o Paulo, Brazil}

\begin{abstract}
In this manuscript we reassess the potential of interacting dark matter-dark energy models in solving the Hubble constant tension. These models have been proposed but also questioned as possible solutions to the $H_0$ problem.
Here we examine several interacting scenarios against cosmological observations, focusing on the important role played by the calibration of Supernovae data.
In order to reassess the ability of interacting dark matter-dark energy scenarios in easing the Hubble constant tension, we systematically confront their theoretical predictions using a prior on the Supernovae Ia absolute magnitude $M_B$, which has been argued to be more robust and certainly less controversial than using a prior on the Hubble constant $H_0$. 
While some data combinations do not show any preference for interacting dark sectors and in some of these scenarios the clustering $\sigma_8$ tension worsens, interacting cosmologies with a dark energy equation of state $w<-1$ are preferred over the canonical $\Lambda$CDM picture even with CMB data alone and also provide values of $\sigma_8$ in perfect agreement with those from weak lensing surveys. Future cosmological surveys will test these exotic dark energy cosmologies by accurately measuring the dark energy equation of state and its putative redshift evolution. 

\end{abstract}



\maketitle

\section{Introduction}
\label{sec:introduction}
A plethora of observations have led to confirm the standard $\Lambda$CDM framework as the most economical and successful model describing our current universe.
This simple picture (pressureless dark matter, baryons and a cosmological constant representing the vacuum energy) has been shown to provide an excellent fit to cosmological data.
However, there are a number of inconsistencies that persist and, instead of diluting with improved precision measurements, gain significance~\cite{Freedman:2017yms,DiValentino:2020zio,DiValentino:2020vvd,DiValentino:2020srs,Freedman:2021ahq,DiValentino:2021izs,Schoneberg:2021qvd,Nunes:2021ipq,Perivolaropoulos:2021jda,Shah:2021onj,Abdalla:2022yfr}.

The most exciting (i.e.\ probably non due to systematics) and most statistically significant ($4-6\sigma$) tension~\cite{Verde:2019ivm,Riess:2019qba,DiValentino:2020vnx} in the literature is the so-called Hubble constant tension, which refers to the discrepancy between cosmological predictions and low redshift estimates of $H_0$.
Within the $\Lambda$CDM scenario, Cosmic Microwave Background (CMB) measurements from the Planck satellite provide a value of $H_0=67.36\pm 0.54$~km s$^{-1}$ Mpc$^{-1}$ at 68\%~CL~\cite{Planck:2018vyg}.
Near universe, local measurements of $H_0$, using the cosmic distance ladder calibration of Type Ia Supernovae with Cepheids, as those carried out by the SH0ES team, provide a measurement of the Hubble constant $H_0=73.04\pm 1.04$~km s$^{-1}$ Mpc$^{-1}$ at 68$\%$~CL~\cite{Riess:2021jrx}.
This problematic $\sim 5\sigma$ discrepancy aggravates when considering other late-time estimates of $H_0$.
For instance, measurements from the Megamaser Cosmology Project~\cite{Pesce:2020xfe}, or those exploiting Surface Brightness Fluctuations~\cite{Blakeslee:2021rqi}, or Type II Supernovae~\cite{deJaeger:2022lit} only exacerbate this tension. Nevertheless, there are measurements from other probes that are still unable to disentangle between the nearby universe and CMB measurements. These include results from the Tip of the Red Giant Branch~\cite{Freedman:2021ahq},
from the astrophysical strong lensing observations~\cite{Birrer:2020tax}
or from gravitational wave events~\cite{Abbott:2017xzu}.
It is important to mention that the differences in $H_0$ measurements between the SHOES and the TRGB approaches arise because the TRGB sample and the Cepheids are modeled as different distance indicators, in many cases belonging to the same host galaxies, see~\cite{Riess:2021jrx} for an updated discussion. Thus, in principle, this difference
may not arise from a different treatment of SNe Ia sample.
On the other hand, several CMB-free analyses (in general, model-dependent analysis) were performed to constraint $H_0$ (see~\cite{Alam_2021,Philcox_2020,Philcox_2022_b,Okamatsu_2021,Nunes_2020_bao,Vagnozzi_age_H0,Abbott_2018,Abbott_2022} for a short list).
Therefore, there is a crucial lack of consensus in the current extraction of the Hubble constant.
The main goal of this manuscript is to scrutinize the ability of some possible late-time non standard cosmologies to solve the issue.

As previously mentioned, the SH0ES collaboration exploits the cosmic distance ladder calibration of Type Ia Supernovae, which means that these observations do not provide a direct extraction of the Hubble parameter.
More concretely, the SH0ES team~\cite{Riess:2020fzl} measures the absolute peak magnitude $M_B$ of Type Ia Supernovae \emph{standard candles} and then translates these measurements into an estimate of $H_0$ by means of the magnitude-redshift relation of the Pantheon Type Ia Supernovae sample~\cite{Scolnic:2017caz}.
Therefore, strictly speaking, the SH0ES team does not directly extract the value of $H_0$, and there have been arguments in the literature aiming to translate the Hubble constant tension into a Type Ia Supernovae absolute magnitude tension $M_B$~\cite{Camarena:2019rmj,Efstathiou:2021ocp,Camarena:2021jlr}.

A number of studies have prescribed to use in the statistical analyses a prior on the intrinsic magnitude rather than on the Hubble constant $H_0$~\cite{Camarena:2021jlr,Schoneberg:2021qvd,Alestas:2020zol,Marra:2021fvf,Alestas:2021luu}.
A prior on $M_B$ is more robust, and avoids double-counting issues when using simultaneously luminosity distance measurements from the SNIa Pantheon sample and measurements of $H_0$ from SH0ES.
In this regard, the value of $H_0$ may also be affected by the choice of the expansion history fit, see Ref.~\cite{Efstathiou:2021ocp}.
We address here in the following the potential of interacting dark matter-dark energy cosmology~\cite{Amendola:1999er} in resolving the Hubble constant tension (\cite{Kumar:2016zpg, Murgia:2016ccp, Kumar:2017dnp, DiValentino:2017iww, Yang:2018ubt, Yang:2018euj, Yang:2019uzo, Kumar:2019wfs, Pan:2019gop, Pan:2019jqh, DiValentino:2019ffd, DiValentino:2019jae, DiValentino:2020leo, DiValentino:2020kpf, Gomez-Valent:2020mqn, Yang:2019uog, Lucca:2020zjb, Martinelli:2019dau, Yang:2020uga, Yao:2020hkw, Pan:2020bur, DiValentino:2020vnx, Yao:2020pji, Amirhashchi:2020qep, Yang:2021hxg, Gao:2021xnk, Lucca:2021dxo, Kumar:2021eev,Yang:2021oxc,Lucca:2021eqy,Halder:2021jiv,Dainotti:2021pqg} and references therein)
by demonstrating explicitly from a full analysis 
that the performance of the model still holds when applying a prior on $M_B$ (see also the recent~\cite{Nunes:2021zzi}).

\section{Theoretical framework}
\label{sec:theory}
We adopt a flat cosmological model described by the Friedmann-Lema\^{i}tre-Robertson-Walker metric.
A possible parameterization of a dark matter-dark energy interaction is provided by the following expressions~\cite{Valiviita:2008iv,Gavela:2009cy}:

\begin{eqnarray}
  \label{eq:conservDM}
\nabla_\mu T^\mu_{(dm)\nu} &=& Q \,u_{\nu}^{(dm)}/a~, \\
  \label{eq:conservDE}
\nabla_\mu T^\mu_{(de)\nu} &=&-Q \,u_{\nu}^{(dm)}/a~.
\end{eqnarray}
In the equations above, $T^\mu_{(dm)\nu}$ and $T^\mu_{(de)\nu}$ represent the energy-momentum tensors for the dark matter and dark energy components respectively, the function $Q$ is the interaction rate between the two dark components, and $u_{\nu}^{(dm)}$ represents the dark matter four-velocity. 
In what follows we shall restrict ourselves to the case in which  the
interaction rate is proportional to the dark energy density $\rho_{de}$~\cite{Valiviita:2008iv,Gavela:2009cy}:
\begin{equation}
Q=\deltadmde\mathcal{H} \rho_{de}~,
\label{rate}
\end{equation}
where $\deltadmde$ is a dimensionless coupling parameter and
$\mathcal{H}=\dot{a}/a$~\footnote{The dot indicates derivative respect to conformal time $d\tau=dt/a$.}.
The background evolution equations in the coupled model considered
here read~\cite{Gavela:2010tm}
\begin{eqnarray}
\label{eq:backDM}
\dot{{\rho}}_{dm}+3{\mathcal H}{\rho}_{dm}
&=&
\deltadmde{\mathcal H}{\rho}_{de}~,
\\
\label{eq:backDE}
\dot{{\rho}}_{de}+3{\mathcal H}(1+\wfld){\rho}_{de}
&=&
-\deltadmde{\mathcal H}{\rho}_{de}~.
\end{eqnarray}

The evolution of the dark matter and dark energy density perturbations and velocities divergence field are described in~\cite{DiValentino:2019jae} and references therein.
We use this modeling to describe the linear perturbation dynamics evolution of the model under consideration in this work. Some of the main effects on the formation of structures in large scales were recently reviewed in detail in~\cite{Nunes_2022_prd}.

It has been shown in the literature that this model is free of instabilities
if the sign of the  coupling $\deltadmde$ and the sign of $(1+\wfld)$ are opposite,
where $\wfld$ refers to the dark energy equation of state~\cite{He:2008si,Gavela:2009cy}.
In order to satisfy such stability conditions, we explore three possible scenarios, all of them with a redshift-independent equation of state.
In Model A,  the equation of state $\wfld$ is fixed to $-0.999$.
Consequently, since $(1+\wfld) >0$, in order to ensure an instability-free perturbation evolution, the dark matter-dark energy coupling $\deltadmde$ is allowed to vary in a negative range.
In Model B, $\wfld$ is allowed to vary but we ensure that the condition $(1+\wfld)>0$ is always satisfied.
Therefore, the coupling parameter $\deltadmde$ is negative.
In Model C, instead, the dark energy equation of state is phantom ($\wfld<-1$), therefore the dark matter-dark energy coupling is taken as positive to avoid early-time instabilities.

The coupling function $Q$ can take different forms.
For instance, it may be proportional to $\rho_{dm}$~\cite{Nunes_2016_rhom,Kumar_2016_rhom,Wang_2016_r} or $\rho_{de}+\rho_{dm}$~\cite{Pan_2020_rewq,Wang_2016_r},
as well as other phenomenological forms~\cite{Yang_2019_uyfh,Pan_2020_mjhg},
or admit a general covariant framework~\cite{Kase_2020_a, Kase_2020_b}.
Admitting a general interaction coupling function and under reasonable  physical stability conditions is possible to restrict the function $Q$ in several ways~\cite{Kase_2020_b}.
In addition, the authors in~\cite{Pan_2020_model_de}, show interaction rates featuring factors proportional to $H$, including but not limited to the case considered here, may naturally emerge when considering well-motivated and simple field theories for the coupling in the dark sector.

As discussed above, we shall present separately the cosmological constraints for three models based on the parameterization eq.~(\ref{rate}), together with those corresponding to the canonical $\Lambda$CDM case.

\begin{table}[t]
    \centering
    \begin{tabular}{c|c|c}
    Model     & Prior $\wfld$ & Prior $\deltadmde$ \\
    \hline
    A     & -0.999 & [-1.0, 0.0]\\
    B     & [-0.999, -0.333] & [-1.0, 0.0] \\
    C     & [-3, -1.001]& [0.0, 1.0] \\
    \end{tabular}
    \caption{Priors of $\wfld$, $\delta$ in models A, B, C.}
    \label{tab:priors}
\end{table}

\section{Datasets and Methodology}
\label{sec:data}

In this Section, we present the data sets and methodology employed to obtain the observational constraints on the model parameters by performing Bayesian Monte Carlo Markov Chain (MCMC) analyses.
In order to constrain the parameters, we use the following data sets:
\begin{itemize}
\item The Cosmic Microwave Background (CMB) temperature and polarization power spectra from the final release of Planck 2018, in particular we adopt the plikTTTEEE+lowl+lowE likelihood~\cite{Aghanim:2018eyx,Aghanim:2019ame}, plus the CMB lensing reconstruction from the four-point correlation function~\cite{Aghanim:2018oex}.
\item Type Ia Supernovae (SN) distance moduli measurements from the \textit{Pantheon} sample~\cite{Scolnic:2017caz}. These measurements constrain the uncalibrated luminosity distance $H_0d_L(z)$ or in other words the slope of the late-time expansion rate (which in turn constrains the current matter energy density, $\Omega_{\rm 0,m}$), where $d_L$ is given by:
\begin{equation}
d_L(z)= c \ (1+z) \int \frac{dz}{H(z)}~.
\end{equation}
\item Baryon Acoustic Oscillations (BAO) distance and expansion rate measurements from the 6dFGS~\cite{Beutler:2011hx}, SDSS-DR7 MGS~\cite{Ross:2014qpa}, BOSS DR12~\cite{Alam:2016hwk} galaxy surveys,
as well as from the eBOSS DR14 Lyman-$\alpha$ (Ly$\alpha$) absorption~\cite{Agathe:2019vsu} and Ly$\alpha$-quasars cross-correlation~\cite{Blomqvist:2019rah}.
These consist of isotropic BAO measurements of $D_V(z)/r_d$
(with $D_V(z)$ and $r_d$ the spherically averaged volume distance and sound horizon at baryon drag, respectively)
for 6dFGS and MGS, and anisotropic BAO measurements of $D_M(z)/r_d$ and $D_H(z)/r_d$
(with $D_M(z)$ the comoving angular diameter distance and $D_H(z)=c/H(z)$ the radial distance)
for BOSS DR12, eBOSS DR14 Ly$\alpha$, and eBOSS DR14 Ly$\alpha$-quasars cross-correlation. 
\item A gaussian prior on $M_B= -19.244 \pm 0.037$~mag~\cite{Camarena:2021jlr}, corresponding to the SN measurements from SH0ES~\cite{Riess:2020fzl}.
When used, this prior is considered over all the Pantheon sample.
\end{itemize}
For the sake of brevity, data combinations are indicated as
CMB (C), CMB+BAO (CB),
CMB+SN (CS),
CMB+SN+BAO (CSB)
and
CMB+SN+BAO+$M_B$ (CSBM).

Cosmological observables are computed with \texttt{CLASS}~\cite{Blas:2011rf,Lesgourgues:2011re}.
In order to derive bounds on the proposed scenarios, we modify the efficient and well-known cosmological package \texttt{MontePython}~\cite{Brinckmann:2018cvx}, supporting the Planck 2018 likelihood~\cite{Planck:2019nip}.
We make use of CalPriorSNIa, a module for \texttt{MontePython}, publicly available at \url{https://github.com/valerio-marra/CalPriorSNIa}, that implements an effective calibration prior on the absolute magnitude of Type Ia Supernovae~\cite{Camarena:2019moy,Camarena:2021jlr}.

\section{Main results and discussion}
\label{sec:results}

\begin{figure*}[t]
\begin{center}
\includegraphics[width=0.7\textwidth]{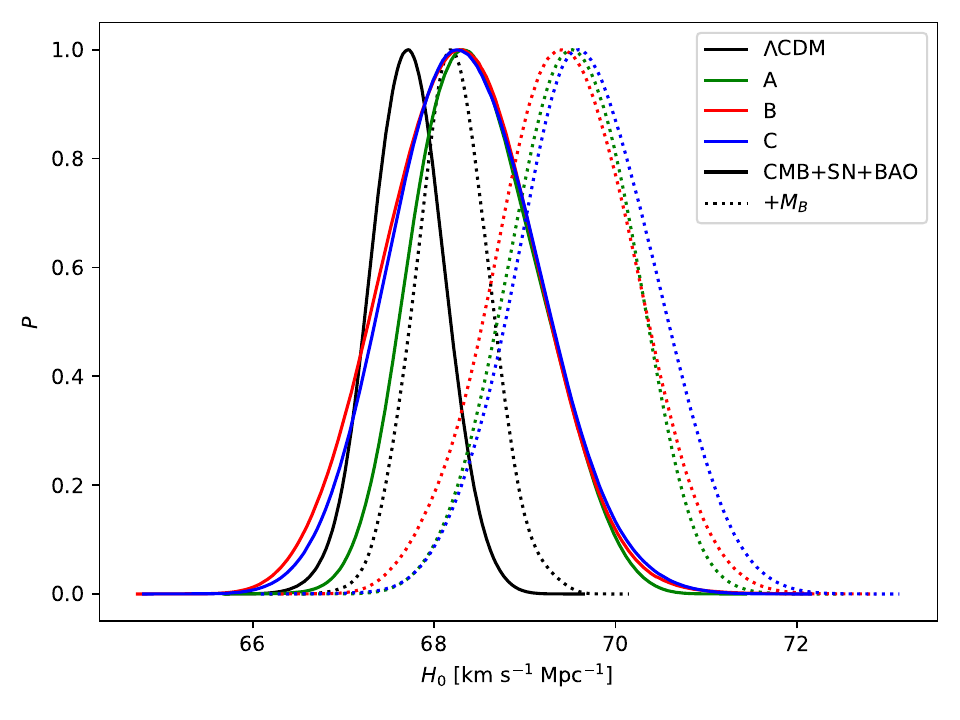} 
\caption{Posterior distribution of the Hubble parameter in the $\Lambda$CDM model (black) and in interacting cosmologies, with priors on the parameters as given in Tab.~\ref{tab:priors}. 
We show constraint obtained within model A (green), model B (red) and model C (blue)
for the CMB+SN+BAO data combination (solid lines)
and CMB+SN+BAO+$M_B$ (dotted lines).
}
\label{fig:h0}
\end{center}
\end{figure*}

\begin{figure*}
\begin{center}
\includegraphics[width=\textwidth]{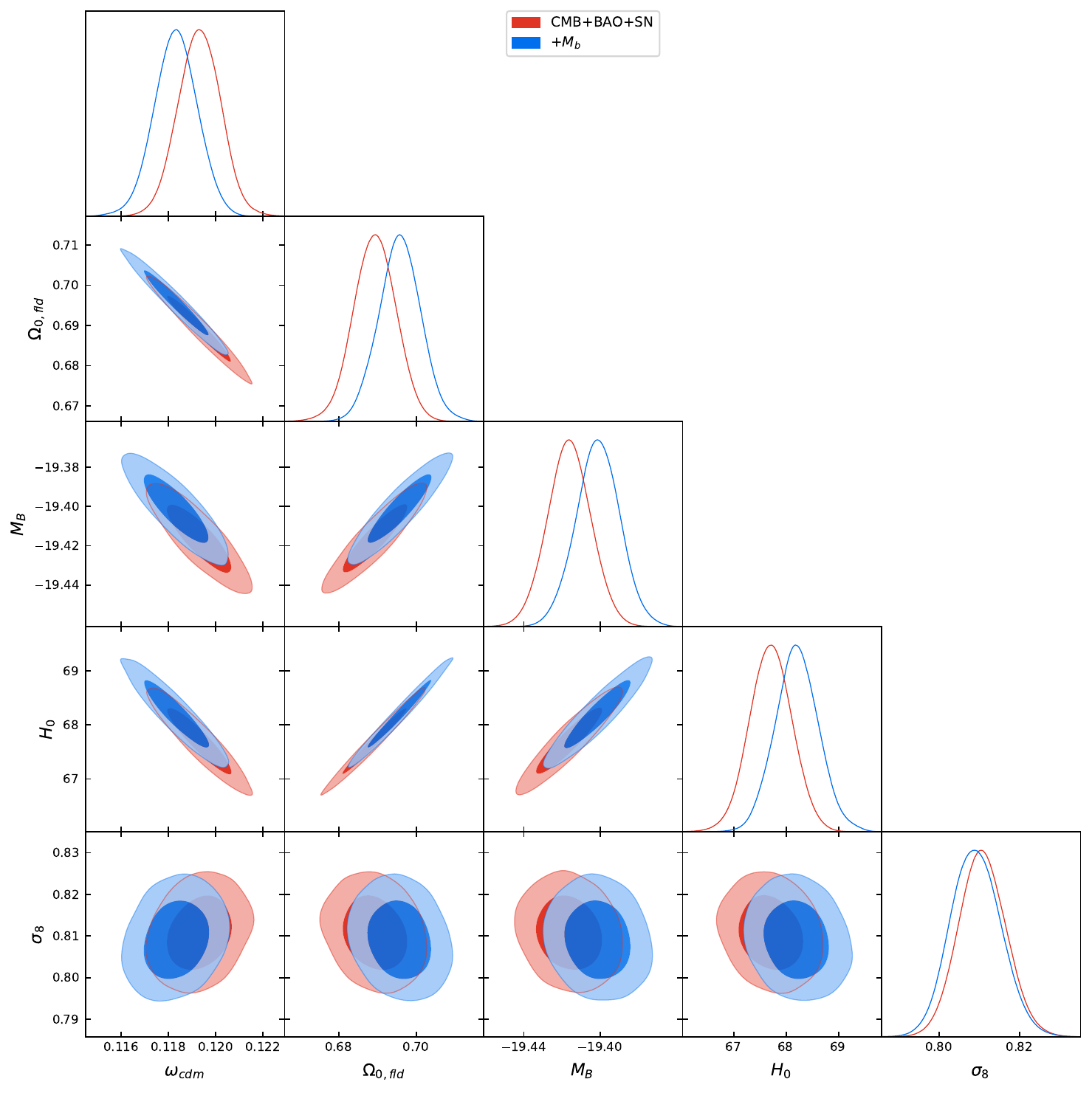} 
\caption{68\% CL and 95\% CL allowed contours and one-dimensional posterior probabilities on a selection of cosmological parameters within the canonical $\Lambda$CDM picture, considering two data combinations: CMB+SN+BAO (red)
and CMB+SN+BAO+$M_B$ (blue).}
\label{fig:triangle_LCDM}
\end{center}
\end{figure*}

\begin{table*}[t]
\centering
\begin{tabular}{|l|c|c|c|c|c|} 
\hline 
Parameter & C & CB & CS & CSB & CSBM \\ \hline
$\omega{}_{cdm }$   & $0.1200\pm0.0012$                       & $0.11948_{-0.00090}^{+0.00092}$         & $0.1198\pm0.0011$                       & $0.11930_{-0.00091}^{+0.00089}$         & $0.11831\pm0.00090$                     \\
$\Omegafld$         & $0.6850_{-0.0077}^{+0.0074}$            & $0.6882_{-0.0056}^{+0.0054}$            & $0.6864_{-0.0070}^{+0.0069}$            & $0.6893\pm0.0054$                       & $0.6957\pm0.0053$                       \\
$M_B$               & -                                       & -                                       & $-19.422\pm0.014$                       & $-19.416\pm0.011$                       & $-19.401\pm0.011$                       \\
$H_0$               & $67.38_{-0.55}^{+0.56}$                 & $67.62_{-0.41}^{+0.42}$                 & $67.50\pm0.51$                          & $67.71\pm0.40$                          & $68.20_{-0.40}^{+0.41}$                 \\
$\sigma_8$          & $0.8114_{-0.0059}^{+0.0057}$            & $0.8106_{-0.0056}^{+0.0058}$            & $0.8113_{-0.0061}^{+0.0063}$            & $0.8107_{-0.0057}^{+0.0060}$            & $0.8090_{-0.0063}^{+0.0060}$            \\

\hline
minimum $\chi^2$   & $2780.70$ & $2792.82$ & $3808.42$ & $3820.76$ & $3840.58$ \\

\hline
\end{tabular}
\caption{Mean values and 68\% CL errors on $\omega_{cdm }\equiv\Omega_{cdm} h^2$, the current dark energy density $\Omegafld$,
the Supernovae Ia intrinsic magnitude $M_B$,
the Hubble constant $H_0$ and the clustering parameter $\sigma_8$ within the standard $\Lambda$CDM paradigm.
We also report the minimum value of the $\chi^2$ function obtained for each of the data combinations.
\label{tab:model_LCDM}
}
\end{table*}

We start by discussing the results obtained within the canonical $\Lambda$CDM scenario.
Table~\ref{tab:model_LCDM} presents the mean values and the $1\sigma$ errors on a number of different cosmological parameters within this standard scenario.
Namely, we show the constraints on
$\omega_{cdm }\equiv\Omega_{0,cdm} h^2$,
the current dark energy density $\Omegafld$,
the Supernovae Ia intrinsic magnitude $M_B$,
the Hubble constant $H_0$ and the clustering parameter $\sigma_8$
arising from the data combinations considered here and above described:
CMB (C),
CMB+BAO (CB),
CMB+SN (CS),
CMB+SN+BAO (CSB)
and
CMB+SN+BAO+$M_B$ (CSBM).

Interestingly, \emph{none} of the parameters barely change their mean values when combining CMB data with additional data sets,
except for the case when we consider the prior on the Supernova Ia absolute magnitude $M_B$.
In this particular  case, the value of $H_0=$ is mildly larger (and, consequently, the values of $\omega_{cdm }$ and $\sigma_8$ mildly smaller),
see the black curves depicted in Fig.~\ref{fig:h0} for the CSBM data combination.
Nevertheless, the parameter ranges agree within $1\sigma$ and therefore within the $\Lambda$CDM picture.
This statement can be clearly confirmed from the contours illustrated in Fig.~\ref{fig:triangle_LCDM}, which presents the two-dimensional allowed contours and the one-dimensional posterior probabilities on the parameters shown in Tab.~\ref{tab:model_LCDM}.
This is because the CSB combination is the leading ingredient in the likelihood, and therefore all the parameters, including $M_B$ and $H_0$, are barely modified with respect to their values within the CSB case, when considering the $M_B$ prior.
Nevertheless, there is an increase in the value of the minimum $\chi^2$ of $\sim 20$ when adding the prior on the SN absolute magnitude.

\begin{table*}[t]
\centering
\begin{tabular}{|l|c|c|c|c|c|} 
\hline 
Parameter & C & CB & CS & CSB & CSBM \\ \hline
$\omega{}_{cdm }$   & $0.075_{-0.027}^{+0.040}$               & $0.1062_{-0.0083}^{+0.0124}$            & $0.1082_{-0.0067}^{+0.0091}$            & $0.1094_{-0.0062}^{+0.0084}$            & $0.0986_{-0.0090}^{+0.0091}$            \\
$\Omegafld$         & $0.806_{-0.092}^{+0.072}$               & $0.726_{-0.034}^{+0.025}$               & $0.720_{-0.026}^{+0.020}$               & $0.717_{-0.023}^{+0.019}$               & $0.748_{-0.024}^{+0.023}$               \\
$\deltadmde$        & $>-0.75$                                & $>-0.36$                                & $-0.109_{-0.043}^{+0.104}$              & $>-0.24$                                & $-0.182_{-0.077}^{+0.080}$              \\
$M_B$               & -                                       & -                                       & $-19.402_{-0.020}^{+0.019}$             & $-19.402_{-0.016}^{+0.015}$             & $-19.377_{-0.014}^{+0.015}$             \\
$H_0$               & $71.18_{-2.97}^{+2.31}$                 & $68.70_{-1.03}^{+0.86}$                 & $68.46_{-0.86}^{+0.78}$                 & $68.45_{-0.72}^{+0.64}$                 & $69.51_{-0.65}^{+0.70}$                 \\
$\sigma_8$          & $1.21_{-0.43}^{+0.36}$                  & $0.907_{-0.096}^{+0.064}$               & $0.893_{-0.071}^{+0.048}$               & $0.881_{-0.064}^{+0.045}$               & $0.963_{-0.085}^{+0.077}$               \\

\hline
minimum $\chi^2$   & $2779.16$ & $2792.78$ & $3807.54$ & $3819.12$ & $3836.00$ \\
$\Delta_{\rm AIC}$ & $0.46$ & $1.96$ & $1.12$ & $0.36$ & $-2.58$ \\

\hline
\end{tabular}
\caption{Mean values and 68\% CL errors or 95\% CL limits on $\omega_{cdm }\equiv\Omega_{cdm} h^2$, the current dark energy density $\Omegafld$, the dimensionless dark matter-dark energy coupling $\deltadmde$, the Supernovae Ia intrinsic magnitude $M_B$, the Hubble constant $H_0$ and the clustering parameter $\sigma_8$ within the interacting model A, see Tab.~\ref{tab:priors}.
We also report the minimum value of the $\chi^2$ function obtained for each of the data combinations and the AIC test with respect to the $\Lambda$CDM case.
\label{tab:model_A}
}
\end{table*}

\begin{figure*}
\begin{center}
\includegraphics[width=\textwidth]{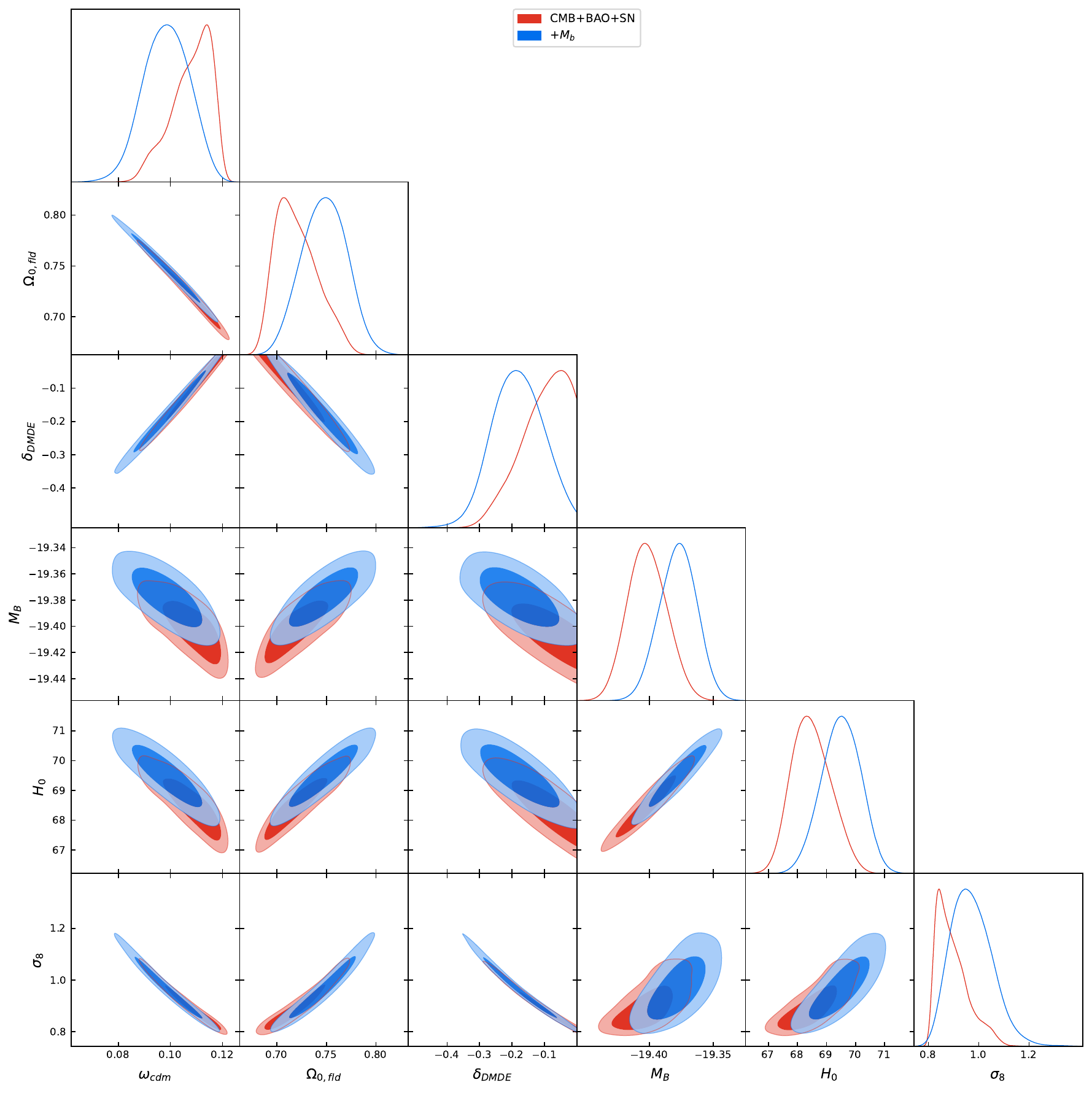} 
\caption{68\% CL and 95\% CL allowed contours and one-dimensional posterior probabilities on a selection of cosmological parameters within model A, considering two data combinations: CMB+SN+BAO (black)
and CMB+SN+BAO+$M_B$ (blue).
}
\label{fig:triangle_A}
\end{center}
\end{figure*}

We focus now on Model A, which refers to an interacting cosmology with $\wfld=-0.999$ and $\deltadmde<0$.
Table~\ref{tab:model_A} presents the mean values and the $1\sigma$ errors on the same cosmological parameters listed above, with the addition of the coupling parameter $\deltadmde$, for the very same data combination already discussed.
Notice that the minimum values of the $\chi^2$ are always smaller than those found for the minimal $\Lambda$CDM scheme for the very same data combinations: therefore, the addition of a coupling \emph{improves} (mildly, in some cases) the overall fit.
Indeed, the $\Delta \chi^2$ between the CSBM and CSB data sets is now $\sim 16$, i.e.\ less than within the $\Lambda$CDM case.
We also quantify here how much the fit improves by considering an information criteria that has been widely exploited in astrophysical and cosmological contexts, namely the frequentist Akaike Information Criterion (AIC)~\cite{Liddle:2007fy,Trotta:2008qt} which establishes that the penalty term between competing models is twice the number of free parameters in the model, and consequently penalizes more complex models that give similar $\chi^2$ values.
The best model is the one minimizing the AIC test: in our tables, positive values prefer $\Lambda$CDM scenarios, while negative ones favor the interacting models.
The significance against a given model will be judged based on the Jeffreys’ scale, which will characterize a difference $\Delta$ AIC $> 5$
($> 10$) as a strong (decisive) evidence against the cosmological model with a higher value for the AIC test. 
Notice from Tab.~\ref{tab:model_A} that data combinations C, CB, CS and CSB show a very mild, almost negligible preference for the $\Lambda$CDM model, while the CSBM data combination prefers a model with a dark sector coupling, leading to the largest $\Delta_{\rm AIC}$ (in absolute value).
This implies that the addition of a dark sector interaction slightly alleviates the tension within the standard $\Lambda$CDM paradigm concerning the inferred values of the Supernova Ia absolute magnitude $M_B$ and that this statement still holds when applying a prior on $M_B$ and not on $H_0$, as the former is considered to be more robust than the latter.
The value of the Hubble constant within the interacting scenario A is larger (regardless of the data set combination) than the one obtained in the $\Lambda$CDM framework (see Fig.~\ref{fig:h0}), especially for the CMB data alone case:
due to the extra contribution from the dark energy component, the amount of intrinsic dark matter should be small and therefore, to leave unchanged the CMB temperature anisotropies peak structure, the value of $H_0$ must be larger.
In addition, one can notice that the value of $\Omegafld$ is much larger.
The reason for this is again related to the lower value for the present matter energy density $\Omega_{\rm 0,m}$, which is required within the interacting cosmologies when the dark matter-dark energy coupling is negative.
In the context of a universe with a negative dark coupling, indeed, there is an energy flow from dark matter to dark energy.
Consequently, the (dark) matter content in the past is higher than in the standard $\Lambda$CDM scenario and the amount of intrinsic (dark) matter needed today is lower, because of the extra contribution from the dark energy sector.
In a flat universe, this translates into a much higher value of $\Omegafld$.

The mean values of the parameters in the CSBM data combination (in which the preferred model is the interacting one) are, in general, in a decent agreement with the mean values obtained within other data set combinations, see Tab.~\ref{tab:model_A}.
Interestingly, we observe a $>2\sigma$ indication in favor of a non-zero value of the coupling $\deltadmde$ when considering  the CSBM data combinations.
Figure~\ref{fig:triangle_A} presents the two-dimensional allowed contours and the one-dimensional posterior probabilities obtained within Model A.
Even if within this interacting scenario there is a $\sim 2\sigma$ indication for a non-zero dark matter-dark energy coupling when considering either $H_0$ or $M_B$ measurements,
the value of the Hubble constant is mildly larger than within the standard $\Lambda$CDM model,
and the model is preferred over the former one applying the AIC criterion to the CSBM data combination (which includes the prior on the SNIa absolute magnitude $M_B$),
the values of the clustering parameter $\sigma_8$ are larger than those within the minimal $\Lambda$CDM picture, increasing the tension with weak lensing surveys.
For instance, the latest results from the Dark Energy Survey~\cite{DES:2021wwk} (see also results for the KiDS-1000 photometric survey~\cite{KiDS:2020suj}) provide $\sigma_8 = 0.733^{+0.039}_{-0.049}$ (assuming a $\Lambda$CDM model) and therefore the addition of a coupling increases the well-known tension between measurements of $\sigma_8$ at high and low redshifts.

\begin{table*}[t]
\centering
\begin{tabular}{|l|c|c|c|c|c|} 
\hline 
Parameter & C & CB & CS & CSB & CSBM \\ \hline
$\omega{}_{cdm }$   & $0.069_{-0.039}^{+0.040}$               & $0.085_{-0.020}^{+0.028}$               & $0.087_{-0.020}^{+0.024}$               & $0.087_{-0.019}^{+0.024}$               & $0.074_{-0.023}^{+0.028}$               \\
$\Omegafld$         & $0.81\pm0.11$                           & $0.768_{-0.063}^{+0.053}$               & $0.764_{-0.056}^{+0.048}$               & $0.765_{-0.055}^{+0.044}$               & $0.800_{-0.061}^{+0.051}$               \\
$\wfld$             & $<-0.73$                                & $<-0.80$                                & $<-0.82$                                & $<-0.81$                                & $<-0.81$                                \\
$\deltadmde$        & $>-0.78$                                & $-0.30_{-0.12}^{+0.27}$                 & $-0.29_{-0.16}^{+0.20}$                 & $-0.28_{-0.15}^{+0.21}$                 & $-0.38_{-0.18}^{+0.21}$                 \\
$M_B$               & -                                       & -                                       & $-19.407_{-0.022}^{+0.023}$             & $-19.405_{-0.018}^{+0.017}$             & $-19.379\pm0.016$                       \\
$H_0$               & $68.72_{-3.40}^{+3.50}$                 & $68.26_{-1.24}^{+1.18}$                 & $68.21_{-1.00}^{+1.02}$                 & $68.28_{-0.82}^{+0.83}$                 & $69.42_{-0.71}^{+0.79}$                 \\
$\sigma_8$          & $1.27_{-0.54}^{+0.51}$                  & $1.07_{-0.27}^{+0.18}$                  & $1.06_{-0.23}^{+0.19}$                  & $1.06_{-0.22}^{+0.17}$                  & $1.21_{-0.34}^{+0.27}$                  \\

\hline
minimum $\chi^2$   & $2778.16$ & $2792.42$ & $3807.22$ & $3818.96$ & $3836.02$ \\
$\Delta_{\rm AIC}$ & $1.46$ & $3.60$ & $2.80$ & $2.20$ & $-0.56$ \\

\hline
\end{tabular}
\caption{Mean values and 68\% CL errors or 95\% CL limits on $\omega_{cdm }\equiv\Omega_{cdm} h^2$, the current dark energy density $\Omegafld$, the dark energy equation of state $\wfld$,
the dimensionless dark matter-dark energy coupling $\deltadmde$, the Supernovae Ia intrinsic magnitude $M_B$, the Hubble constant $H_0$ and the clustering parameter $\sigma_8$ within the interacting model B, see Tab.~\ref{tab:priors}.
We also report the minimum value of the $\chi^2$ function obtained for each of the data combinations and the AIC test with respect to the $\Lambda$CDM case.}
\label{tab:model_B}
\end{table*}

\begin{figure*}
\begin{center}
\includegraphics[width=\textwidth]{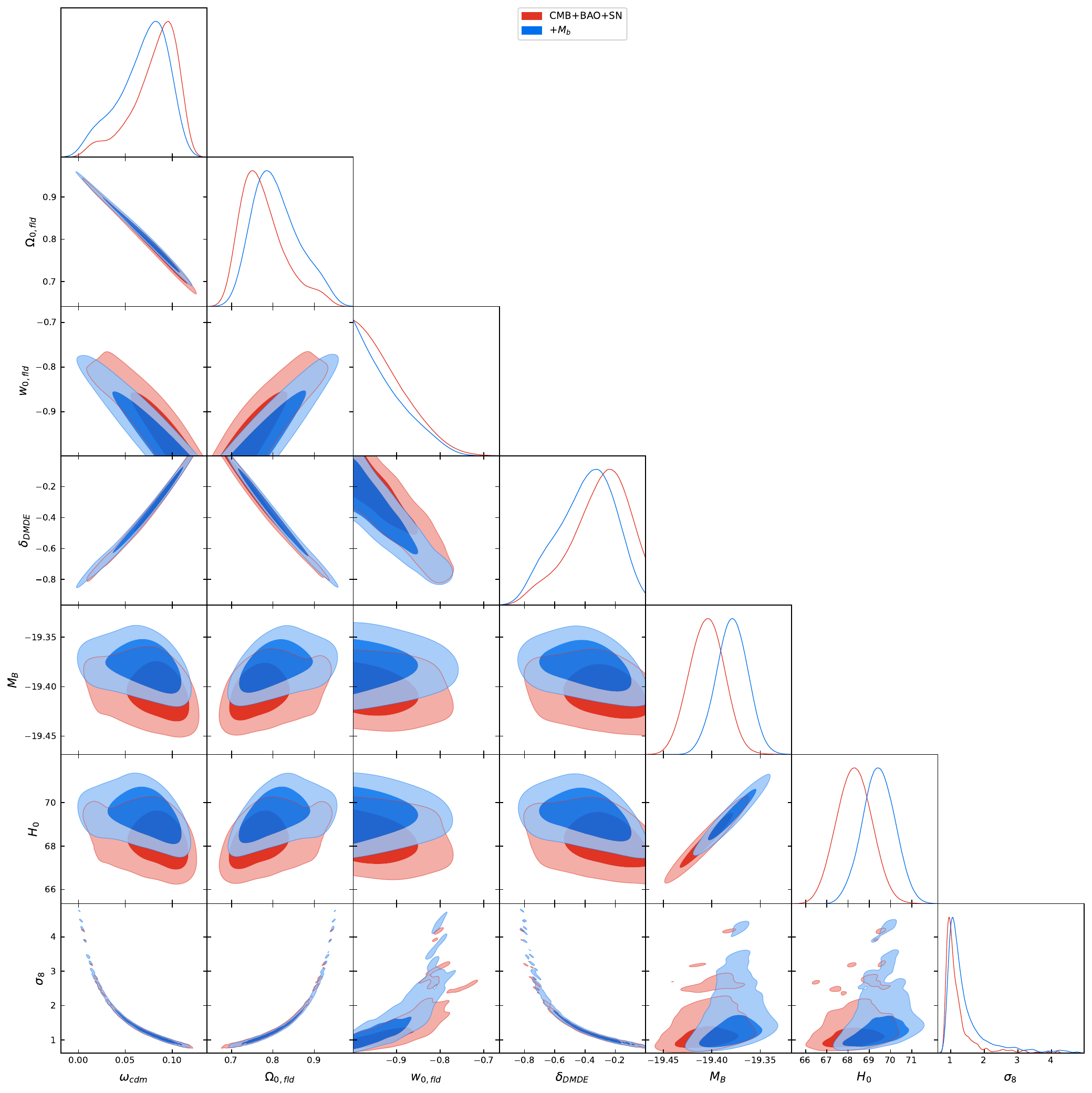} 
\caption{68\% CL and 95\% CL allowed contours and one-dimensional posterior probabilities on a selection of cosmological parameters within model B, considering two data combinations: CMB+SN+BAO (red)
and CMB+SN+BAO+$M_B$ (blue).
}
\label{fig:triangle_B}
\end{center}
\end{figure*}

Focusing now on Model B (see Tab.~\ref{tab:model_B}), which assumes a negative coupling $\deltadmde$ and a constant, but freely varying, dark energy equation of state $\wfld$ within the $\wfld>-1$ region,
we notice, as in Model A, a smaller value of the matter density and a larger value of the Hubble parameter (see Fig.~\ref{fig:h0}) to compensate for the effect of the energy flow between the dark energy and dark matter sectors.
The value of $H_0$, however, is slightly smaller than in Model A, due to the strong anti-correlation between $\wfld$ and $H_0$~\cite{DiValentino:2016hlg,DiValentino:2019jae}, since a larger value of $\wfld>-1$ implies a lower value of $H_0$.
Nevertheless, a mild preference (up to $\sim 2\sigma$) for a non-zero value of the dark matter-dark energy coupling is present also in this case,
for four out of the five data combinations presented in this study.

Notice that the minimum $\chi^2$ in Model B is smaller than that corresponding to the minimal $\Lambda$CDM framework, and also smaller than that of Model A, which is nested in Model B, as expected.
However there is no indication for a preference of this Model B over the $\Lambda$CDM canonical one from the C, CB, CS ans CSB data sets, as can be inferred from the values of $\Delta_{\rm AIC}$ shown in Tab.~\ref{tab:model_B}.
Moreover, even if $\Delta_{\rm AIC}<0$ for the CSBM case, its value is very small and therefore not statistically significant.

In Fig.~\ref{fig:triangle_B} we depict the two-dimensional allowed contours and the one-dimensional posterior probabilities obtained for Model B.
From a comparison to Fig.~\ref{fig:triangle_LCDM} and also confronting the mean values of Tab.~\ref{tab:model_B} to those shown in Tab.~\ref{tab:model_LCDM} (and, to a minor extent, to those in Tab.~\ref{tab:model_A}), one can notice that the value of $\Omegafld$ is again much larger, due to the lower value of $\Omega_{m,0}$.
On the other hand, a lower value of $\Omega_{m,0}$ requires a larger value of the clustering parameter $\sigma_8$ to be able to satisfy the overall normalization of the matter power spectrum, exacerbating the $\sigma_8$ tension with DES~\cite{DES:2021wwk} and KiDS-1000~\cite{KiDS:2020suj} results.

\begin{table*}[t]
\centering
\begin{tabular}{|l|c|c|c|c|c|} 
\hline
Parameter  & C & CB & CS & CSB & CSBM \\ \hline
$\omega{}_{cdm }$   & $0.158_{-0.028}^{+0.024}$               & $0.158_{-0.021}^{+0.019}$               & $0.157_{-0.023}^{+0.021}$               & $0.150\pm0.019$                         & $0.147_{-0.017}^{+0.015}$               \\
$\Omegafld$         & $0.769_{-0.051}^{+0.055}$               & $0.615_{-0.043}^{+0.047}$               & $0.615_{-0.054}^{+0.050}$               & $0.629_{-0.045}^{+0.041}$               & $0.650_{-0.032}^{+0.035}$               \\
$\wfld$             & $-1.81_{-0.23}^{+0.22}$                 & $-1.171_{-0.085}^{+0.109}$              & $-1.168_{-0.085}^{+0.109}$              & $-1.133_{-0.070}^{+0.089}$              & $-1.158_{-0.060}^{+0.064}$              \\
$\deltadmde$        & $<0.77$                                 & $0.41_{-0.31}^{+0.21}$                  & unconstrained                           & $<0.71$                                 & $0.28_{-0.22}^{+0.13}$                  \\
$M_B$               & -                                       & -                                       & $-19.404_{-0.024}^{+0.022}$             & $-19.405\pm0.018$                       & $-19.375_{-0.016}^{+0.017}$             \\
$H_0$               & $>74.12$                                & $68.58_{-1.41}^{+1.30}$                 & $68.40_{-1.07}^{+0.99}$                 & $68.32_{-0.78}^{+0.83}$                 & $69.66_{-0.73}^{+0.79}$                 \\
$\sigma_8$          & $0.75_{-0.12}^{+0.11}$                  & $0.657_{-0.075}^{+0.066}$               & $0.661_{-0.083}^{+0.071}$               & $0.682_{-0.074}^{+0.067}$               & $0.702_{-0.063}^{+0.059}$               \\

\hline
minimum $\chi^2$   & $2773.16$ & $2792.32$ & $3805.90$ & $3819.16$ & $3834.68$ \\
$\Delta_{\rm AIC}$ & $-3.54$ & $3.50$ & $1.48$ & $2.40$ & $-1.90$ \\

\hline
\end{tabular}
\caption{Mean values and 68\% CL errors or 95\% CL limits on $\omega_{cdm }\equiv\Omega_{cdm} h^2$, the current dark energy density $\Omegafld$, the current matter energy density $\Omega_{\rm 0,m}$, the dark energy equation of state $\wfld$,
the dimensionless dark matter-dark energy coupling $\deltadmde$, the Supernovae Ia intrinsic magnitude $M_B$, the Hubble constant $H_0$ and the clustering parameter $\sigma_8$ within the interacting model C, see Tab.~\ref{tab:priors}.
We also report the minimum value of the $\chi^2$ function obtained for each of the data combinations and the AIC test with respect to the $\Lambda$CDM case.}
\label{tab:model_C}
\end{table*}

\begin{figure*}
\begin{center}
\includegraphics[width=\textwidth]{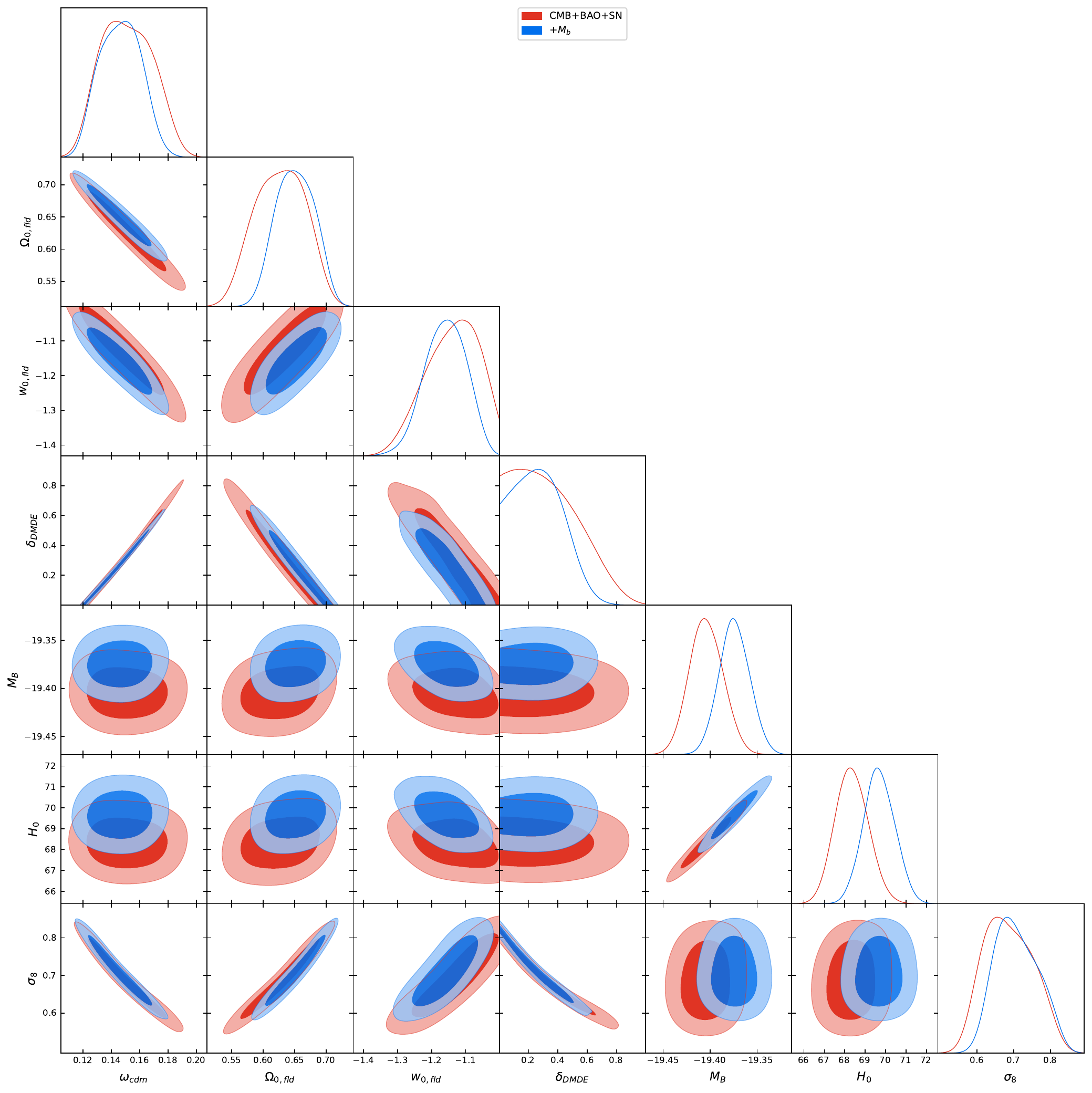} 
\caption{68\% CL and 95\% CL allowed contours and one-dimensional posterior probabilities on a selection of cosmological parameters within model C, considering two data combinations: CMB+SN+BAO (red)
and CMB+SN+BAO+$M_B$ (blue).
}
\label{fig:triangle_C}
\end{center}
\end{figure*}

Finally, Tab.~\ref{tab:model_C} shows the mean values and the $1\sigma$ errors on the usual cosmological parameters explored along this study, for Model C.
Notice that this model benefits from both its interacting nature and from the fact that $\wfld<-1$ and $\deltadmde>0$.
Both features of the dark energy sector have been shown to be excellent solutions to the Hubble constant problem.
Indeed, within this model, the value of the Hubble constant is naturally larger than within the $\Lambda$CDM model (see the blue lines in Fig.~\ref{fig:h0}),
regardless of the data sets assumed in the analyses.
Despite its phantom nature, as in this particular case $\wfld<-1$ to ensure a instability-free evolution of perturbations, Model C provides the \emph{best-fits to any of the data combinations explored here, performing even better than} the minimal $\Lambda$CDM picture,
as one can clearly notice from the results of Tab.~\ref{tab:model_C}.
In addition, the AIC criterion shows that not only the CSBM data combination, but also CMB data alone do prefer this model over the minimal $\Lambda$CDM picture.

Figure~\ref{fig:triangle_C} illustrates the two-dimensional allowed contours and the one-dimensional posterior probabilities obtained within Model C.
Notice that here the situation is just the opposite one of Model B: the value of $\Omegafld$ is much smaller than in standard scenarios, due to the larger value required for the present matter energy density $\Omega_{\rm 0,m}$ when the dark matter-dark energy coupling $\deltadmde>0$ and $\wfld<-1$.
This larger value of the present matter energy density also implies a lower value for the clustering parameter $\sigma_8$, in contrast to what was required within Models A and B, alleviating the $\sigma_8$ tension.
These two facts (i.e.\ best fit model to any of the data combinations plus the alleviation of the $\sigma_8$ clustering parameter tension) make Model C a very attractive cosmological scenario which can provide a solution for the long-standing $H_0$ tension.
Nevertheless, we must remember here that Model C has two degrees of freedom more than the standard $\Lambda$CDM paradigm.

\section{Final Remarks}
\label{sec:conclusions}

In this study we have tried to reassess the ability of interacting dark matter-dark energy cosmologies in alleviating the long-standing and highly significant Hubble constant tension.
Despite the fact that in the past these models have been shown to provide an excellent solution to the discrepancy between local measurements and high redshift (Cosmic Microwave Background) estimates of $H_0$, there have been recent works in the literature questioning their effectiveness, related to a misinterpretation of SH0ES data, which indeed does not directly extract the value of $H_0$.
We have therefore computed the ability of interacting cosmologies of reducing the Hubble tension by means of a prior on Type Ia Supernova absolute magnitude, which is more robust and avoids double counting issues when it is combined with Type Ia Supernovae (SN) luminosity distance observations. 
We combine this prior with Cosmic Microwave Background (CMB), Type Ia Supernovae (SN) and Baryon Acoustic Oscillation (BAO) measurements, showing that despite the value of the Hubble parameter is larger, in general, there is no significant preference for interacting dark energy, and in some cosmologies the well-known $\sigma_8$ tension worsens.
There is however one among the interacting cosmologies considered here,
with a phantom nature,
which provides a better fit than the canonical $\Lambda$CDM framework for all the considered data combinations, is preferred over the canonical $\Lambda$CDM picture from CMB data alone and alleviates the $\sigma_8$ problem.
Nevertheless, this model has two extra degrees of freedom, both describing exotic dark energy physics.
Future galaxy surveys will be able to further test these non-standard dark energy cosmologies by accurately extracting the value of the dark energy equation of state and its possible redshift evolution.

\begin{acknowledgments}
\noindent  
SG acknowledges financial support from the European Union's Horizon 2020 research and innovation programme under the Marie Skłodowska-Curie grant agreement No 754496 (project FELLINI).
EDV is supported by a Royal Society Dorothy Hodgkin Research Fellowship. OM is supported by the Spanish grants PID2020-113644GB-I00, PROMETEO/2019/083 and by the European ITN project HIDDeN (H2020-MSCA-ITN-2019//860881-HIDDeN). RCN acknowledges financial support from the Funda\c{c}\~{a}o de Amparo \`{a} Pesquisa do Estado de S\~{a}o Paulo (FAPESP, S\~{a}o Paulo Research Foundation) under the project No. 2018/18036-5.
\end{acknowledgments}

\bibliography{PRD}

\end{document}